\documentclass[11pt]{article}
\usepackage{multirow}
\usepackage{graphicx} 
\usepackage{amssymb}
\usepackage{amsmath}
\usepackage{bigints}
\usepackage{color}
\usepackage{multirow}
\usepackage{color}
\usepackage[colorlinks,linkcolor=blue,citecolor=green]{hyperref}
\usepackage[top=3cm,bottom=3cm,left=2.5cm,right=2.5cm]{geometry}

\newtheorem{theorem}{Theorem}

\newenvironment{proof}[1][Proof]{\textbf{#1.} }{\ \rule{0.5em}{0.5em}}

\usepackage{graphicx}
\allowdisplaybreaks[3]

\title{\textbf{Classification of 1$+$0 two-dimensional Hamiltonian operators}\\[10mm]}
\author{Alessandra Rizzo
\\[5mm]
 \small  Department of Mathematics\\
\small  University of Palermo, \\
\small  \texttt{alessandra.rizzo07@unipa.it} }

\date{}
\begin{document}

\maketitle

\begin{abstract}
In this paper, we study Hamiltonian operators which are sum of a first-order one and a Poisson tensor, in two spatial independent variables. In particular, a complete classification of these operators is presented in two and three components, analyzing both the cases of degenerate and non degenerate leading coefficients.

\end{abstract}
\section{Introduction}
The Hamiltonian formalism represents a connection between geometry and mathematical physics, playing also a fundamental role in the analysis of nonlinear systems. A wide number of examples of systems possessing the Hamiltonian property arise from physics, biology, engineering and applied mathematics in general \cite{Maj, River, River2}. A building block in this theory is represented by quasilinear systems of PDEs, i.e. systems of hydrodynamic type, given by equations of the form
\begin{equation*}
	u_{t}^{i}= v_{l}^{i \alpha}(\textbf{u})\dfrac{\partial u^{l}}{\partial x^\alpha}, \qquad i,l =1,\dots n,   \qquad \alpha =1,\dots N.
\end{equation*}
Here the Einstein's convention on repeated indices is used. In the previous formula, we considered $x^1,\dots x^N$ the independent spatial variables, $t$ the evolution independent variable, $u^1,\dots u^n$ the field variables depending both on $x^1,\dots , x^N$ and $t$. In this case, we say that the system is an evolutionary one in $N$ dimensions, or a system in $N+1$ independent variables. 

The Hamiltonian formalism for hydrodynamic type systems was deeply investigated in the last fifty years \cite{DN83,DN01,Getz, Sciacca}. In analogy with the finite-dimensional case, i.e. for Ordinary Differential Equations (also known as OdE), the Hamiltonian structure is strictly related to the existence of Poisson brackets, that is skew-symmetric brackets satisfying the Jacobi identity. Indeed, an evolutionary system $u^i_t=F^i$ is said Hamiltonian with respect to the Poisson brackets $\{,\, ,\, \}$ if there exists a function (or more generally a functional) $H$ such that
\begin{equation}
	u^i_t=F^i=\{u^i,H\}, \qquad i=1,2,\dots n.
\end{equation}

In the one dimensional case (obtained for $ \alpha=1$), a Poisson bracket is called of hydrodynamic type if it takes the form
\begin{equation*}
	\{u^i(x), u^j (y) \} =g^{i j}(\textbf{u}(x)) \delta' (x-y)+ b_{k}^{ij } (\textbf{u}(x)) u_{x}^{k}(x) \delta(x-y),
\end{equation*}
where $u = (u^1, u^2, \dots, u^n)$ are local coordinates on a smooth $n$-dimensional manifold, $x$ and $y$ are independent variables, $u_x^k$ is the partial derivative of the coordinate $u^k$ with respect to the variable $x$, $g$ and $b$ are coefficients depending only on the field variables and $\delta$ is the Dirac delta-function. These brackets were firstly studied by Dubrovin and Novikov \cite{DN83} and after that many contributions to different generalisations of the theory of Poisson brackets of hydrodynamic type were given     \cite{Doyle, Femo, Femo2,Pote, Pote2,PavVerVit} .\\
In particular, starting from the existence of important physical examples of hydrodynamic type systems in two spatial dimensions (gas dynamic and shallow water equations,  combustion theory, general relativity, nonlinear elasticity and so on {\cite{Olver,DN001, Ts11}),
	Dubrovin and Novikov  \cite{DN83,DubrovinNovikov:PBHT} presented the problem of the classification of multidimensional Poisson brackets of hydrodynamic type, i.e. Poisson brackets of the form
	\begin{equation}
		\{u^i(x), u^j (y) \} =\sum_{\alpha=1}^{n} g^{i j\alpha}(u(x)) \delta_\alpha (x-y)+ b_{k}^{ij \alpha} (u(x)) u_{\alpha}^{k}(x) \delta(x-y).
		\label{equazionedelleparentesi}
	\end{equation}

	In \eqref{equazionedelleparentesi},  $u = (u^1, u^2, \dots, u^n)$ are local coordinates on a smooth $n$-dimensional manifold, $x=(x^1, x^2, \dots x^N)$ and $y=(y^1, y^2, \dots, y^N)$ are independent variables, $u_\alpha^k$ is the partial derivative of the coordinate $u^k$ with respect to the variable $x^\alpha$, $g^{ij \alpha}$ and $b_{k}^{ij \alpha}$ are coefficients depending only on the field variables. Furthermore $\delta_\alpha(x-y)=\dfrac{\partial \delta (x-y)}{\partial x^\alpha}$ and $\delta$ is the Dirac delta-function.
	
	Let us now recall that given a Poisson bracket $\{\, ,\,\}$, it is possible to define an integro-differential operator $P^{ij}$ such that
	for each pair of functionals $f$ and $g$ one has
	\begin{equation}\label{bra}
		\{f, g \}= \int \dfrac{\delta  f}{\delta u^{i}} P^{ij} \dfrac{\delta  g}{\delta u^{j}}\, dx,
	\end{equation}
	
	where $\delta/\delta u^k$ is the variational derivative of its argument with respect to the $k$th field variable.
	
	An operator $P$ associated to a Poisson bracket is a Hamiltonian operator, i.e. it is a skew-adjoint differential operator with vanishing Schouten torsion ( that is $[P, P] = 0$). Notice that the converse is also true, so that if $P^{ij}$ is a Hamiltonian operator, the associated brackets defined as in \eqref{bra} are Poisson brackets. 
	\\
	In this framework, the differential operator $P^{ij}$ associated to \eqref{equazionedelleparentesi} takes the form
	\begin{equation}
		P^{ij}(u)=g^{ij \alpha} \dfrac{d}{d x^\alpha}+ b_{k}^{ij \alpha} u_{\alpha}^{k}
		\label{operator}
	\end{equation} 
	and it is defined so that 
	$P^{ij}$ is Hamiltonian and it is the so called \emph{multidimensional hydrodynamic type operator}.  \\
	When dealing with these operators, a fundamental distinction is necessary: the degenerate and the non degenerate case. 
	We recall that the operator \eqref{operator} is non degenerate if, $\forall \, \alpha$ , $\det(g^{i j \alpha} )\neq 0$. In the non degenerate case,  Dubrovin and Novikov proposed the problem to find necessary and sufficient conditions on the coefficients $g^{ij\alpha}$ and $b^{ij\alpha}_k$ such that the operator $P^{ij}(u)$ is Hamiltonian. We remark that in the one dimensional case, the solution was given by the authors themselves who proved that $P^{ij}$ is Hamiltonian if and only if $g_{ij}=(g^{ij})^{-1}$ is a flat metric whose Christoffel symbols are related to $b^{ij}_k$ as $b^{ij}_k=-g^{is}\Gamma^{j}_{sk}$. The final answer was given  by Mokhov \cite{Mk1,mok}, who provided the conditions that guarantee the Hamiltonianity of multidimensional Poisson brackets of hydrodynamic type for any number $n$ of components and for any dimension $N$ using the theory of compatible
	metrics constructed by the same author in \cite{Mk2}.
	On the other hand, the degenerate case was studied by Savoldi \textit{et al.} in \cite{
		Savoldi1, Savoldi2}, where a complete classification of two and three components Hamiltonian operators is given.\\
	Furthermore, Dell'Atti and Vergallo \cite{DeVe} analysed non-homogeneous Hamiltonian operators composed of a first order Dubrovin-Novikov operator and an ultralocal one, providing a full classification of 1D two and three components operators of the form 
	\begin{equation}\label{10}
		P^{ij}+ \omega^{ij}= \left(g^{ij} \dfrac{d}{d x}+ b_{k}^{ij} u^{k}_x\right)\quad +\quad \omega^{ij},
	\end{equation}
	that play a key role in  the integrability theory of non-homogeneous quasilinear systems of first order PDEs \cite{Ver2, Ver3}. Operators \eqref{10} are known as \emph{non-homogeneous hydrodynamic operators} or $1+0$ operators. An important example of a system possessing a Hamiltonian structure through this kind of operators is given by the three-wave equation
	\begin{equation*}
		\begin{cases}
			u_t=-a\,u_x-2\,(b-c) \, v\, w\\
			v_t=-b\,v_x-2\,(a-c) \, u\, w\\
			w_t=-c\,w_x-2\,(b-a) \, u\, v
		\end{cases},
	\end{equation*}
	that is Hamiltonian with
	\begin{equation*}
		P^{ij}+ \omega^{ij}= \begin{pmatrix}
			1 & 0 &0\\
			0 &-1&0\\
			0 & 0 &-1
		\end{pmatrix} \dfrac{\partial}{\partial x}+ \begin{pmatrix}
			0 & - 2 \, w & 2 \, v\\
			2 \, w &0 &2 \, u\\
			-2 \,v & - 2 \, u &0
		\end{pmatrix}.
	\end{equation*}
	Recently, the work of  Dell'Atti and Vergallo has been slightly extended by Hu and Casati \cite{Casati} (who indeed investigated in details the necessary conditions for non-homogeneous systems to be Hamiltonian with the same type of operators here considered), providing a classification of the operators of the form
	\begin{equation}
		P^{ij}(u)+ \omega^{ij}=g^{ij \alpha} \dfrac{d}{d x^\alpha}+ b_{k}^{ij \alpha} u_{\alpha}^{k}+\omega^{ij},
		\label{operator2}
	\end{equation}
	in the two dimensional non-degenerate case with two components.\\
	In the same paper, the authors also prove for the first time that the two dimensional three-wave equation
	\begin{equation*}
		\begin{cases}
			u_t=-a\,u_x+  d\, u_y-2\,(b-c) \, v\, w\\
			v_t=-b\,v_x+ e\, v_y-2\,(a-c) \, u\, w\\
			w_t=-c\,w_x+ f w_y-2\,(b-a) \, u\, v
		\end{cases}, \: \: \: \: \: \: \: \: f=\dfrac{d(c-b)+ e(a-c)}{a-b},
	\end{equation*}
	is Hamiltonian with a multidimensional non-homogeneous hydrodynamic operator
	\begin{equation*}
		P^{ij}+ \omega^{ij}= \begin{pmatrix}
			S & 0 &0 \\
			0 & S &0\\
			0&0& S
		\end{pmatrix} \dfrac{\partial}{ \partial x}+   \begin{pmatrix}
			\frac{d-e}{a-b} \, S & 0 &0 \\
			0 &   \frac{d-e}{a-b} \, S &0\\
			0&0&   \frac{d-e}{a-b} \, S
		\end{pmatrix} \dfrac{\partial}{ \partial y}+  \begin{pmatrix}
			0 & S \, w &-S \, v \\
			-S \, w &  0 &S \, u\\
			S \, v&-S \, u&  0
		\end{pmatrix},
	\end{equation*}
	where $S$ is an arbitrary constant.\\
	
	Motivated by this viewpoint, the aim of our work is to make a step forward in this analysis: we want to extend the analysis performed by Casati $et \, al.$ to the degenerate two components case and to present a novel classification of the operators of form $(\ref{operator2})$ in $2D$ ($\alpha=1, \,2$) and three components ($i, \, j , \, k= 1, \, 2,\,3$). In Section \ref{2}, we discuss the conditions that guarantee the Hamiltonianity of the multi-dimensional operators under interest. In Section \ref{3}, we present a complete classification for the two component case. Finally, in Section \ref{4} the three components case is treated.
	After that, conclusions and some final remarks are given.

	\section{Hamiltonianity of multidimensional non-homogeneous hydrodynamic operators }
	In this section, we briefly review the conditions for the operator \eqref{operator2} to be Hamiltonian, i.e. the conditions that guarantee that the bracket 
	\begin{equation}\{f, g\}  = \int \dfrac{\delta f }{\delta u^{i}} (P^{ij}+ \omega^{ij}) \dfrac{\delta g }{\delta u^{j}} d x^1 d x^2 \dots d x^n \end{equation}
	satisfies the property of skew-symmetry ($\{f,g\}=-\{g,f\}$) and the Jacobi identity ($\{\{f,g\},h\}+\{\{g,h\},f\}+\{\{h,f\},g\}=0$).
	
	\vspace{10mm}

	We notice that the operator under interest is given by the first-order homogeneous differential operator
	\begin{equation}
		P^{ij}=g^{ij \alpha} \dfrac{d}{d x^\alpha}+ b_{k}^{ij \alpha} u_{\alpha}^{k}
		\label{ophom}
	\end{equation}
	\label{2}
	and by the ultralocal operator $w^{ij}$ of order $0$. In addition, being the Schouten torsion a bilinear bracket, every operator which is non-homogeneous in the degree of derivation, but sum of two homogeneous ones, at least requires both the addenda to be Hamiltonian. In our case, this implies that both $P^{ij}$ and $\omega^{ij}$ must satisfy their Hamiltonianity conditions respectively. Furthermore, a supplementary compatibility condition involving $P^{ij}$ and $\omega^{ij}$ must hold.\\
	Let us start with the lower dimensional operator, i.e. let us firstly focus on the necessary and sufficient conditions for $\omega^{ij}$ to be Hamiltonian. We recall the following Theorem:\\ 
	
	\begin{theorem}[\cite{mokhov98}]
		\label{hamw}
		The ultralocal operator $w^{ij}$ is Hamiltonian if and only if it satisfies
		\begin{align*}
			&\omega^{ij}=-\omega^{ij},\\
			&\omega^{is} \dfrac{\partial \omega^{jk}}{\partial u^{s}}+\omega^{js} \dfrac{\partial \omega^{ki}}{\partial u^{s}}+\omega^{ks} \dfrac{\partial \omega^{ij}}{\partial u^{s}}=0.
		\end{align*}
	\end{theorem}
	We observe that in the non degenerate case, the above conditions are respectively
	skew-symmetry and closedness of the 2-form $\alpha=(\omega)^{-1}$.

	On the other hand, for multidimensional hydrodynamic operator, the following statement holds true: \\
	\begin{theorem}[\cite{Mk1}]
		\label{hamP}
		The operator $P^{ij}$ is Hamiltonian if and only if the following conditions are satisfied
		\begin{align*}
			&g^{i j \alpha}=g^{j i \alpha},\\
			&\dfrac{\partial g^{i j \alpha}}{\partial u^k}=b_{k}^{ij \alpha}+b_{k}^{j i \alpha},\\
			&\sum_{(\alpha, \, \beta)} (g^{li \alpha} b_{l}^{j k \beta}-g^{l j \beta} b_{l}^{i k \alpha})=0,\\
			&\sum_{(i, \, j, \, k)} (g^{li \alpha} b_{l}^{j k \beta}-g^{l j \beta} b_{l}^{i k \alpha})=0,\\
			&\sum_{(\alpha, \, \beta)} 
			\left[ g^{li \alpha} \left( \dfrac{\partial b_{l}^{j k \beta} }{\partial u^{r}}- \dfrac{\partial b_{r}^{j k \beta} }{\partial u^{l}}\right)+  b_{l}^{i j  \alpha}  b_{r}^{l k \beta}- b_{l}^{i k\alpha}  b_{r}^{l j \beta}\right]=0,\\
			&g^{l i \beta} \dfrac{\partial b_{r}^{jk \alpha}}{\partial u^{l}}-g^{l  j \alpha} \dfrac{\partial b_{r}^{ i k \beta}}{\partial u^{l}}- b_{l}^{i j  \beta}  b_{r}^{l k \alpha}- b_{l}^{i k \alpha}  b_{r}^{l j \beta}+ b_{l}^{ j i   \alpha}  b_{r}^{l k \beta}+b_{l}^{j k \alpha}  b_{r}^{i l \beta}=0,\\
			& \dfrac{\partial}{\partial u^{s}}\left[ g^{li \alpha} \left( \dfrac{\partial b_{l}^{j k \beta} }{\partial u^{r}}- \dfrac{\partial b_{r}^{j k \beta} }{\partial u^{l}}\right)+  b_{l}^{i j  \alpha}  b_{r}^{l k \beta}- b_{l}^{i k\alpha}  b_{r}^{l j \beta}\right]+\\
			& \dfrac{\partial}{\partial u^{r}}\left[ g^{li \beta} \left( \dfrac{\partial b_{l}^{j k \alpha} }{\partial u^{s}}- \dfrac{\partial b_{s}^{j k \alpha} }{\partial u^{l}}\right)+  b_{l}^{i j  \beta}  b_{s}^{l k \alpha}- b_{l}^{i k \beta}  b_{s}^{l j \alpha}\right]+\\
			& \sum_{(i, \, j , \, k)} \left[ b_{r}^{l i \beta} \left(\dfrac{\partial b_{s}^{j k \alpha} }{\partial u^{l}}- \dfrac{\partial b_{l}^{j k \alpha} }{\partial u^{s}} \right) \right]+ \sum_{(i, \, j , \, k)} \left[ b_{s}^{l i \alpha} \left(\dfrac{\partial b_{r}^{j k \beta} }{\partial u^{l}}- \dfrac{\partial b_{l}^{j k \beta} }{\partial u^{r}} \right) \right]=0,
		\end{align*}
	\end{theorem}
	where $\sum$ represents the cyclic summation over the displayed 
	indexes. Furthermore, in the above, $i,j,k,l,r= 1, \dots n$ and $\alpha, \, \beta=1 , \dots N.$\vspace{0.25cm}\\
	
	We notice that in the case of $N=1$ with non degenerate leading coefficient, this Theorem reduces to the Dubrovin and Novikov's result \cite{DN83}.\\
	
	At this point, we need only to remind the following theorem stating the conditions for the non homogeneous operator \eqref{operator2}.\\
	\begin{theorem}[\cite{Casati, mokhov98}]
		\label{hamtu}
		The non homogeneous multidimensional operator \eqref{operator2} is Hamiltonian if and only if
		\begin{itemize}
						\item $w^{ij}$ satisfies the conditions given in Theorem \ref{hamw};
						\item $P$ satisfies the conditions given in Theorem \ref{hamP} ;
			\item given $T^{ijk \alpha}= g^{il \alpha} w_{l}^{jk}-b_{l}^{ij\alpha} w^{lk}-b_{l}^{ik\alpha} w^{jl}$, the supplementary conditions
			\begin{align}
				&T^{ijk\alpha}=T^{kij \alpha} \label{condizioneT}\\
				&T_{s}^{ijk \alpha}=b_{s}^{lk \alpha} w_{l}^{ij}+ (b_{s,l}^{ki \alpha}- b_{l,s}^{ki \alpha}) w^{lj}\label{condizioneTder} 
			\end{align}
			are verified.\end{itemize}
	\end{theorem}

	\section{Classification in $2D$ and two components}
	\label{3}
	In this paper, we provide a classification of the $1+0$ degenerate operators of form \eqref{operator2} in $2D$ and in the case of two and three components. As pointed out before, we recall that for the homogeneous case of the first order operator \eqref{operator}, a full classification in $2D$ with two and three components has been proposed by Mokhov \cite{mok} in the non degenerate case and by  Savoldi $et \, al.$ in \cite{Savoldi1, Savoldi2} in the degenerate case. Hence, the problem of classifying Hamiltonian operators of form \eqref{operator2} is reduced to the problem of classifying their ultralocal part and its compatibility with the first order homogeneous Hamiltonian operators already classified in literature. In order to do so, we need to require that the conditions given in Theorems \ref{hamw} and \ref{hamtu} are verified.\\
	The needed calculations have been performed with the use of computer algebra methods. The importance of symbolic computation for the analysis of Hamiltonian structures is itself a topic of research \cite{cpal1, cpal2}.\\
	In this section, we focus on the two components case, with $u$ and $v$ depending on $t$ and on the variables $x$ and $y$. Casati and Hu \cite{Casati} already provided the classification for the non degenerate case, so for this number of components we treat only the degenerate one. 
	
	We consider the two possible forms of the degenerate homogeneous operator provided by Savoldi in \cite{Savoldi1}, i.e.
	\begin{equation}
		P_1= \begin{pmatrix}
			1 &0\\
			0&0
		\end{pmatrix} \dfrac{d}{d x}+ \begin{pmatrix}
			v &0 \\
			0 &0
		\end{pmatrix} \dfrac{d}{dy}+ \begin{pmatrix}
			\frac{1}{2} v_y &0 \\
			0 &0
		\end{pmatrix}
	\end{equation}
	and
	\begin{equation}
		P_2=\begin{pmatrix}
			1 &0\\
			0&0
		\end{pmatrix}\dfrac{d}{d x}+ \begin{pmatrix}
			v &0 \\
			0 &0
		\end{pmatrix}\dfrac{d}{dy}+ \begin{pmatrix}
			\frac{1}{2} v_y &- \frac{v_x+ v v_y}{u}  \vspace{0.2cm}\\
			\frac{v_x+ v v_y}{u} &0.
		\end{pmatrix}
	\end{equation} and the ultralocal structure 
	\begin{equation}
		\omega= \begin{pmatrix}
			0 & f(u,v)\\
			-f(u,v) &0
		\end{pmatrix}.
	\end{equation}
	At this point, we look for the form of non-homogeneous structures starting from the
	pairs $(P_i, \omega)$. Taking into account theorems \ref{hamw}, and \ref{hamtu}, we are able to give the following\\ 
	\begin{theorem}
		The operator $P_1 + \omega$ is Hamiltonian if and only if $f(u,v)= F(v)$, where $F(v)$ is an arbitrary function.  The operator $P_2 + \omega$ is Hamiltonian if and only if $f(u,v)= \frac{F(v)}{u}$, where $F(v)$ is an arbitrary function.  
	\end{theorem}
\begin{proof}
The first and the second condition of Theorem $\ref{hamtu}$ are trivially satisfied by the operators under consideration. Hence, we just need to compute conditions \eqref{condizioneT} and \eqref{condizioneTder}. Here, we explicit the calculations for $P_1+\omega.$\\ Fixing $i=k=\alpha=1$ and $j=2,$ we obtain
\begin{equation*}
T^{1211}-T^{1121}=	2 \dfrac{\partial f(u,v)}{\partial u}=0,
\end{equation*}
from which we derive $f(u,v)=F(v).$ At this point, it is simple matter to verify that \eqref{condizioneT} and \eqref{condizioneTder} are identically satisfied. 
\end{proof}

	\section{Classification in $2D$ and three components}
	\label{4}
	In this section, we focus on the case of three components $u, \, v, \, w$ depending on the two variables $x$ and $y$. Once again, we start from the classification of the homogeneous operators and we investigate the possible compatible ultralocal structure. We conduct our analysis in both the degenerate and the non degenerate case. 
	\subsection{Non degenerate case}
	We refer to the classification provided by Ferapontov, Lorenzoni and Savoldi \cite{Savoldi2}. They proved that any non degenerate and non-constant three-component Hamiltonian operator in $2D$ can
	be brought (by a change of the dependent variables $u, \, v$ and $w$) into one of the
	two following canonical forms 
	\begin{align*}
		&P_3= \begin{pmatrix}
			0 & 0 &1\\
			0 & 1 &0\\
			1 & 0 &0
		\end{pmatrix} \dfrac{d}{dx}+\begin{pmatrix}
			-2 v & w &\lambda\\
			w & \lambda &0\\
			\lambda & 0 &0  
		\end{pmatrix}\dfrac{d}{dy}+\begin{pmatrix}
			-v_y & 2 w_y &0\\
			-w_y & 0 &0\\
			0 & 0 & 0
		\end{pmatrix},\\ \vspace{0.3cm}\\
		&P_4=\begin{pmatrix}
			0 & 0 &1\\
			0 & 1 &0\\
			1 & 0 &0
		\end{pmatrix} \dfrac{d}{dx}+\begin{pmatrix}
			-2 u & -\frac{1}{2}v &w \vspace{0.1cm}\\
			-\frac{1}{2}v & w &0 \vspace{0.1cm}\\
			w & 0 &0  
		\end{pmatrix}\dfrac{d}{dy}+\begin{pmatrix}
			-u_y & \frac{1}{2} v_y & 2 w_y \vspace{0.1cm}\\
			-v_y & \frac{1}{2} w_y &0\vspace{0.1cm}\\
			-w_y & 0 & 0
		\end{pmatrix}.  
	\end{align*}
	We consider the ultralocal structure
	\begin{equation}
		\omega= \begin{pmatrix}
			0 & f_1(u,v,w) & f_2(u,v,w)\\
			-f_1(u,v,w) &  0 & f_3(u,v,w)\\
			-f_2(u,v,w) & -f_3(u,v,w) &0
		\end{pmatrix}
		\label{forma}
	\end{equation}
	and we impose the conditions for the operators $P_3+ \omega$ and $P_4+ \omega$ to be Hamiltonian.\\
	Computing the requirements resulting from theorems \ref{hamw} and \ref{hamtu}, it follows\\
	\begin{theorem}
		The operator $P_3 + \omega$ is Hamiltonian if and only if $f_1=c \, u+c_1$, $f_2=-c \, v$ and $f_3=c \, w$, where $c$ and $c_1$ are arbitrary constants. The operator $P_4+ \omega$ is Hamiltonian if and only if    $f_1=f_2=f_3=0$.
		\end{theorem}
	\subsection{Degenerate case}
	Here, we classify the $1+0$  Hamiltonian operators in the degenerate case. We start from the classification  of the homogeneous first order Hamiltonian operators provided by Savoldi in \cite{Savoldi1}. We keep in mind that degenerate $2D$ structures of hydrodynamic type, defined through a pair
	of metrics $g^{ij1}$ and $g^{ij 2}$, are classified according to the rank of the pencil $g_\lambda=g^{ij1}-\lambda g^{ij2}.$ For this reason, in the following, we distinguish the cases of rank equal to 0, 1 or 2. \vspace{0.2cm}\\
	\textbf{Rank 0}\\
	In the case $ rank(g_\lambda)=0$, any non constant three component Hamiltonian operator of Dubrovin-Novikov type in dimension two can be brought into one of the two following forms
	\begin{align*}
		& P_5=\begin{pmatrix}
			0 &w_x+u w_y &0\\
			-w_x-u w_y & 0 & 0\\
			0 &0 &0
		\end{pmatrix},\\  
		& P_6=\begin{pmatrix}
			0 &w_x+w w_y &0\\
			-w_x-w w_y & 0 & 0\\
			0 &0 &0
		\end{pmatrix}.
	\end{align*}
	We consider the ultralocal structure $\omega$ of the form \eqref{forma}. Imposing the conditions for $P_5+ \omega$ and $P_6+\omega$ to be Hamiltonian, we find \\
	\begin{theorem}
		The operators $P_5+ \omega$ and $P_6+ \omega$ are Hamiltonian if and only if $f_2=f_3=0$, while $f_1$ is an arbitrary function of the variables $u, \, v$ and $w$.
	\end{theorem}
\vspace{0.2cm}
	\textbf{Rank 1}\\
	In the case $rank(g_\lambda)=1$, there are five possible  canonical forms of non-trivial degenerate three component Hamiltonian operators
	of Dubrovin-Novikov type. For each, we consider an ultralocal structure of form \eqref{forma} and we study the Hamiltonianity. \vspace{0.2cm}\\
	-The operator $P_7+\omega$, where $P_7$ is
	\begin{equation}
		P_7=\begin{pmatrix}
			1 &0 &0\\
			0 & 0 & 0\\
			0 &0 &0
		\end{pmatrix} \dfrac{d }{dx}+\begin{pmatrix}
			0 &0 &h(v,w) v_y\\
			0 & 0 & 0\\
			-h(v,w) v_y &0 &0
		\end{pmatrix}
	\end{equation}
	is Hamiltonian if and only if 
	\begin{enumerate}
		\item $h=h(v,w)$, $f_1=f_3=0$ and $f_2=f_2(v,w)$, where $h$ and $f_2$ are arbitrary functions of the variables $v$ and $w$;
		\item $h=f_3=0$, $f_1=f_1(v,w)$ and $f_2=f_2(v,w)$, where $f_1$ and $f_2$ are arbitrary functions of the variables $v$ and $w$
		\item $h=0$, $f_1=f_1(v,w)$, $f_3=f_3(v,w)$, where $f_1$ and $f_3$ are arbitrary functions of the variables $v$ and $w$, while
		\begin{equation}
			f_2= -\left[\int \dfrac{\partial }{\partial v} \left(\dfrac{f_1(v,w)}{f_3(v,w)} \right)\,dw +f_4(v) \right] f_3(v,w),
		\end{equation}
		with $f_4(v)$ is an arbitrary function. 
	\end{enumerate}
	\vspace{0.2cm}
	- The operator $P_8+\omega$, where $P_8$ is
	\begin{equation}
		P_8=\begin{pmatrix}
			1 &0 &0\\
			0 & 0 & 0\\
			0 &0 &0
		\end{pmatrix} \dfrac{d }{dx}+\begin{pmatrix}
			v &0 &0\\
			0 & 0 & 0\\
			0 &0 &0
		\end{pmatrix} \dfrac{d }{dy}+\begin{pmatrix}
			\dfrac{1}{2}v_y &0 &h(v,w) v_y \\
			0 & 0 & 0\\
			-h(v,w) v_y &0 &0
		\end{pmatrix}
	\end{equation}
	is Hamiltonian if and only if 
	\begin{enumerate}
		\item $h=h(v,w)$, $f_1=f_3=0$ and $f_2=f_2(v,w)$. where $h$ and $f_2$ are arbitrary functions of the variables $v$ and $w$;
		\item $h=f_3=0$, $f_1=f_1(v,w)$ and $f_2=f_2(v,w)$, where $f_1$ and $f_2$ are arbitrary functions of the variables $v$ and $w$.
	\end{enumerate}
	- The operator $P_9+\omega$, where $P_9$ is
	\begin{equation}
		P_9=\begin{pmatrix}
			1 &0 &0\\
			0 & 0 & 0\\
			0 &0 &0
		\end{pmatrix} \dfrac{d }{dx}+\begin{pmatrix}
			f(v,w) &0 &0\\
			0 & 0 & 0\\
			0 &0 &0
		\end{pmatrix} \dfrac{d }{dx}+\begin{pmatrix}
			\dfrac{f_v v_y+ f_w w_y}{2} &w_x+h(v,w) w_y &0 \vspace{0.1cm}\\
			-w_x-h(v,w) w_y  & 0 & 0\\
			0&0 &0
		\end{pmatrix}
	\end{equation}
	is Hamiltonian if and only if $f_2=f_3=0$ and $f_1=f_1(v,w)$, where $f_1$ is an arbitrary function of the variables $v$ and $w$.   \vspace{0.2cm} \\
	- The operator $P_{10}+\omega$, where $P_{10}$ is
	\begin{align*}
		-P_{10}=&\begin{pmatrix}
			1 &0 &0\\
			0 & 0 & 0\\
			0 &0 &0
		\end{pmatrix} \dfrac{d }{dx}+\begin{pmatrix}
			f(v,w) &0 &0\\
			0 & 0 & 0\\
			0 &0 &0
		\end{pmatrix} \dfrac{d }{dx}+ \vspace{0.2cm}\\ & \, \, \, \, \,\begin{pmatrix}
			\dfrac{f_v v_y+ f_w w_y}{2} &0 &-\dfrac{w_x+f(v,w) w_y-h(v,w) v_y}{u} \vspace{0.2cm}\\
			0   & 0 & 0\\
			\dfrac{w_x+f(v,w) w_y-h(v,w) v_y}{u}&0 &0
		\end{pmatrix}
	\end{align*}
	is Hamiltonian if and only if $f_1=f_3=0$ and $f_2(u,v,w)=\dfrac{F_2(v,w)}{u}$, with $F_2$ arbitrary function  of the variables $v$ and $w$.   \vspace{0.2cm}\\
	-     The operator $P_{11}+\omega$, where $P_{11}$ is
	\begin{align*}
		P_{11}=\begin{pmatrix}
			1 &0 &0\\
			0 & 0 & 0\\
			0 &0 &0
		\end{pmatrix} \dfrac{d }{dx}+\begin{pmatrix}
			v &0 &0\\
			0 & 0 & 0\\
			0 &0 &0
		\end{pmatrix} \dfrac{d }{dy}+\begin{pmatrix}
			\dfrac{ v_y}{2} &-\dfrac{v_x+v v_y}{u} &-\dfrac{w_x+v w_y}{u} \vspace{0.1cm}\\
			\dfrac{v_x+v v_y}{u}   & 0 & 0 \vspace{0.1cm}\\
			\dfrac{w_x+v w_y}{u}&0 &0
		\end{pmatrix}
	\end{align*}
	is Hamiltonian if and only if $f_3=0$, $f_1= \dfrac{F_1(v,w)}{u}$ and  $f_2= \dfrac{F_2(v,w)}{u}$, with $F_1$ and $F_2$ arbitrary functions  of the variables $v$ and $w$. \vspace{0.2cm}\\
	\textbf{Rank 2}\\
	As a last case, we consider $rank(g_\lambda)=2.$ For each canonical form, we consider an ultralocal structure of form \eqref{forma} and we give the conditions for the operator $P_i+\omega$ to be Hamiltonian. \vspace{0.2cm}\\
	- The operator $P_{12}+ \omega$, where $P_{12}$ is
	\begin{equation*}
		P_{12}=\begin{pmatrix}
			0 &1 &0\\
			1 & 0 & 0\\
			0 &0 &0
		\end{pmatrix} \dfrac{d }{dx}+\begin{pmatrix}
			-2 u &v &0\\
			v & 0 & 0\\
			0 &0 &0
		\end{pmatrix}\dfrac{d }{dy}+\begin{pmatrix}
			-u_y &2 v_y &0\\
			-v_y& 0 & 0\\
			0 &0 &0
		\end{pmatrix} 
	\end{equation*}
	is Hamiltonian if and only if $f_1=f_2=f_3=0$.
	\vspace{0.2cm}\\
	-The operator $P_{13}+ \omega$, where $P_{13}$ is
	\begin{equation*}
		P_{13}=\begin{pmatrix}
			0 &1 &0\\
			1 & 0 & 0\\
			0 &0 &0
		\end{pmatrix} \dfrac{d }{dx}+\begin{pmatrix}
			-2 u &v &0\\
			v & 0 & 0\\
			0 &0 &0
		\end{pmatrix}\dfrac{d }{dy}+\begin{pmatrix}
			-u_y &2 v_y &w_y\\
			-v_y& 0 & 0\\
			-w_y &0 &0
		\end{pmatrix} 
	\end{equation*}
	is Hamiltonian if and only if $f_1=f_3=0$, while $f_2$ is an arbitrary function of the variable $w$.\vspace{0.2cm}\\
	-The operator $P_{14}+ \omega$, where $P_{14}$ is
	\begin{equation*}
		P_{14}=\begin{pmatrix}
			0 &1 &0\\
			1 & 0 & 0\\
			0 &0 &0
		\end{pmatrix} \dfrac{d }{dx}+\begin{pmatrix}
			0 &0 &1\\
			0 & 0 & 0\\
			1 &0 &0
		\end{pmatrix} \dfrac{d }{dy}
	\end{equation*}
	is Hamiltonian if and only if $f_1$, $f_2$ and $f_3$ are constants. \vspace{0.2cm}\\
	-The operator $P_{15}+ \omega$, where $P_{15}$ is
	\begin{equation*}
		P_{15}=\begin{pmatrix}
			0 &1 &0\\
			1 & 0 & 0\\
			0 &0 &0
		\end{pmatrix} \dfrac{d }{dx}+\begin{pmatrix}
			p(w) &q(w) &0\\
			q(w) & r(w) & 0\\
			0 &0 &0
		\end{pmatrix}\dfrac{d }{dy}+\begin{pmatrix}
			\dfrac{p'}{2} w_y &0 &0\\
			q'(w) w_y & \dfrac{r'}{2}w_y & 0\\
			0 &0 &0
		\end{pmatrix} 
	\end{equation*}
	is Hamiltonian if and only if 
	\begin{enumerate}
		\item  $f_1=f_1(w)$, $f_2=f_2(w)$ and $f_3=c f_2(w)$ (with  $f_1$ and $f_2$ arbitrary functions of the variable $w$), in the case in which $p(w)=c_1$, $r(w)=2 c q(w)+c_2$ and $q(w)$ is an arbitrary function. In the previous relations, $c$. $c_1$ and $c_2$ are arbitrary constants;
		\item $f_1=f_1(w)$, $f_3=f_3(w)$   (with  $f_1$ and $f_3$ arbitrary functions of the variable $w$) and $f_2=0$ if $p$, $q$ are arbitrary constants, while $r$ is an arbitrary function of $w$; 
		\item  $f_1=f_1(w)$, $f_2=f_2(w)$   (with  $f_1$ and $f_2$ arbitrary functions of the variable $w$) and $f_3=0$ if $p$, $q$ are arbitrary functions of $w$, and $r$ is equal to a constant ;
		\item  $f_1=f_1(w)$ (with  $f_1$ arbitrary function of the variable $w$), $f_2=f_3=0$ ,otherwise.
	\end{enumerate} \vspace{0.2cm}
	-The operator $P_{16}+ \omega$, where $P_{16}$ is
	\begin{equation*}
		P_{16}=\begin{pmatrix}
			0 &1 &0\\
			1 & 0 & 0\\
			0 &0 &0
		\end{pmatrix} \dfrac{d }{dx}+\begin{pmatrix}
			p(w) &q(w) &0\\
			q(w) & r(w) & 0\\
			0 &0 &0
		\end{pmatrix}\dfrac{d }{dy}+\begin{pmatrix}
			\frac{p'}{2} w_y &w_y &0\vspace{0.1cm}\\
			(q'(w) -1)w_y & \frac{r'}{2}w_y & 0\vspace{0.1cm}\\
			0 &0 &0
		\end{pmatrix} 
	\end{equation*}
	is Hamiltonian if and only if 
	\begin{enumerate}
		\item  $f_1=f_1(w)$, $f_2=f_2(w)$ and $f_3=c f_2(w)$ (with $c$ constant,  $f_1$ and $f_2$ arbitrary functions of the variable $w$), in the case in which $p=\dfrac{2 w}{c}+c_1$, $q$ is an arbitrary function of $w$ and $r=2 c (q-w)+c_2$. In the previous relations, $c, c_1$ and $c_2$ are arbitrary constants ;
		\item  $f_1=f_1(w)$, $f_3=f_3(w)$ (with  $f_1$ and $f_3$ arbitrary functions of the variable $w$) and $f_2=0$ if $r$ is an arbitrary functions of $w$, $p$ is constant and $q= c+ w$, where $c$ is equal to a constant;
		\item  $f_1=f_1(w)$ (with  $f_1$ arbitrary function of the variable $w$) and $f_2=f_3=0$ otherwise.
	\end{enumerate} \vspace{0.2cm}
	-The operator $P_{17}+ \omega$, where $P_{17}$ is
	\begin{equation*}
		P_{17}=\begin{pmatrix}
			0 &1 &0\\
			1 & 0 & 0\\
			0 &0 &0
		\end{pmatrix} \dfrac{d }{dx}+\begin{pmatrix}
			1 &0 &0\\
			0 & 0 & 0\\
			0 &0 &0
		\end{pmatrix}\dfrac{d }{dy}+\begin{pmatrix}
			0 &0 &-\dfrac{w_x}{v}\\
			0 & 0 & 0\\
			\dfrac{w_x}{v} &0 &0
		\end{pmatrix} 
	\end{equation*}
	is Hamiltonian if and only if $f_3=0$, $f_1=f_1(w)$ and $f_2=\dfrac{F_2(w)}{v}$ where $f_1$ and $F_2$ are arbitrary functions of the variable $w$. \vspace{0.2cm}\\
	-The operator $P_{18}+ \omega$, where $P_{18}$ is
	\begin{equation*}
		P_{18}=\begin{pmatrix}
			0 &1 &0\\
			1 & 0 & 0\\
			0 &0 &0
		\end{pmatrix} \dfrac{d }{dx}+\begin{pmatrix}
			0 &w &0\\
			w & 0 & 0\\
			0 &0 &0
		\end{pmatrix}\dfrac{d }{dy}+\begin{pmatrix}
			0 &0 &-\dfrac{w_x+w w_y}{v}\vspace{0.1cm}\\
			w_y & 0 & 0\vspace{0.1cm}\\
			\dfrac{w_x+w w_y}{v} &0 &0
		\end{pmatrix} 
	\end{equation*}
	is Hamiltonian if and only if
	$f_3=0$, $f_1=f_1(w)$ and $f_2=\dfrac{F_2(w)}{v}$, with $f_1$ and $F_2$ are arbitrary functions of the variable $w$. \vspace{0.2cm}\\
	-The operator $P_{19}+ \omega$, where $P_{19}$ is
	\begin{equation*}
		P_{19}=\begin{pmatrix}
			0 &1 &0\\
			1 & 0 & 0\\
			0 &0 &0
		\end{pmatrix} \dfrac{d }{dx}+\begin{pmatrix}
			1 &w &0\\
			w & 0 & 0\\
			0 &0 &0
		\end{pmatrix}\dfrac{d }{dy}+\begin{pmatrix}
			0 &0 &-\dfrac{w_x+w w_y}{v}\vspace{0.1cm}\\
			w_y & 0 & 0\vspace{0.1cm}\\
			\dfrac{w_x+w w_y}{v} &0 &0
		\end{pmatrix} 
	\end{equation*}
	is Hamiltonian if and only if 
	$f_3=0$, $f_1=f_1(w)$ and $f_2=\dfrac{F_2(w)}{v}$, with $f_1$ and $F_2$ arbitrary functions. \vspace{0.2cm}\\
	-The operator $P_{20}+ \omega$, where $P_{20}$ is
	\begin{equation*}
		P_{20}=\begin{pmatrix}
			0 &1 &0\\
			1 & 0 & 0\\
			0 &0 &0
		\end{pmatrix} \dfrac{d }{dx}+\begin{pmatrix}
			0 &0 &0\\
			0 & 0 & 1\\
			0&1 &0
		\end{pmatrix}\dfrac{d }{dy}+\begin{pmatrix}
			0 &0 &-\dfrac{w_x-u_y}{v}\\
			0 & 0 & 0\\
			\dfrac{w_x-u_y}{v} &0 &0
		\end{pmatrix} 
	\end{equation*}
	is Hamiltonian if and only if $f_1=f_3=0$ and $f_2=\dfrac{c}{v}$, where $c$ is a constant. \vspace{0.2cm}\\
	-The operator $P_{21}+ \omega$, where $P_{21}$ is
	\begin{equation*}
		P_{21}=\begin{pmatrix}
			0 &1 &0\\
			1 & 0 & 0\\
			0 &0 &0
		\end{pmatrix} \dfrac{d }{dx}+\begin{pmatrix}
			0 &0 &1\\
			0 & 0 & 0\\
			1&0 &0
		\end{pmatrix}\dfrac{d }{dy}+\begin{pmatrix}
			0 &0 &-\dfrac{w_x-v_y}{v}\\
			0 & 0 & 0\\
			\dfrac{w_x-v_y}{v} &0 &0
		\end{pmatrix} 
	\end{equation*}
	is Hamiltonian if and only if $f_1=c$, $f_3=0$ and $f_2=\dfrac{c \, w+ c_1}{v}$ where $c$ and $c_1$ are constants. \vspace{0.2cm}\\
	- The operator $P_{22}+ \omega$, where $P_{22}$ is
	\begin{equation*}
		P_{22}=\begin{pmatrix}
			0 &1 &0\\
			1 & 0 & 0\\
			0 &0 &0
		\end{pmatrix} \dfrac{d }{dx}+\begin{pmatrix}
			u & -\dfrac{v}{2} &0 \\
			- \dfrac{v}{2} & 0 & 0\\
			0&0 &0
		\end{pmatrix}\dfrac{d }{dy}+\begin{pmatrix}
			\dfrac{u_y}{2} &-v_y &-\dfrac{w_x}{v} \vspace{0.1cm}\\
			\dfrac{v_y}{2}& 0 & 0 \vspace{0.1cm}\\
			\dfrac{w_x}{v} &0 &0
		\end{pmatrix} 
	\end{equation*}
	is Hamiltonian if and only if $f_1=f_3=0$ and $f_2=\dfrac{F_2( w)}{v}$ where $F_2$ is an arbitrary function of the variable $w$. \vspace{0.2cm}\\
	-The operator $P_{23}+ \omega$, where $P_{23}$ is
	\begin{equation*}
		P_{23}=\begin{pmatrix}
			0 &1 &0\\
			1 & 0 & 0\\
			0 &0 &0
		\end{pmatrix} \dfrac{d }{dx}+\begin{pmatrix}
			1& -w &0 \\
			- w & w^2 & 0\\
			0&0 &0
		\end{pmatrix}\dfrac{d }{dy}+\begin{pmatrix}
			0 &0 &\dfrac{w_x-2 w w_y}{wu-v}\\
			-w_y& w w_y & -\dfrac{w w_x-2 w^2 w_y}{wu-v}\\
			-\dfrac{w_x-2 w w_y}{wu-v}&\dfrac{w w_x-2 w^2 w_y}{wu-v} &0
		\end{pmatrix} 
	\end{equation*}
	is Hamiltonian if and only if $f_1=F_1(w)$ and $f_2=\dfrac{-2 w F_1( w)}{wu-v}$ and $f_3=\dfrac{2 w^2 F_1( w)}{wu-v}$  where $F_1(w)$ is an arbitrary function. \vspace{0.2cm}\\
	-The operator $P_{24}+ \omega$, where $P_{24}$ is
	\begin{equation*}
		P_{24}=\begin{pmatrix}
			0 &1 &0\\
			1 & 0 & 0\\
			0 &0 &0
		\end{pmatrix} \dfrac{d }{dx}+\begin{pmatrix}
			1& -k &0 \\
			- k & k\, w & 0\\
			0&0 &0
		\end{pmatrix}\dfrac{d }{dy}+\begin{pmatrix}
			-\dfrac{k w_y}{w^2} &\dfrac{k w_y}{2 w} &\dfrac{w_x-2 k w_y}{wu-v} \vspace{0.1cm}\\
			-\dfrac{k w_y}{2 w}& \dfrac{w_y}{2} & -\dfrac{w w_x-2 k w w_y}{wu-v} \vspace{0.1cm}\\
			-\dfrac{w_x-2 k w_y}{wu-v}&\dfrac{w w_x-2 k w w_y}{wu-v} &0
		\end{pmatrix} 
	\end{equation*}
	is Hamiltonian if and only if $f_1=f_2=f_3=0$.

	\section{Conclusions}
	Non-homogeneous operators of order $k + m$ (i. e. operators composed by the sum of two homogeneous operators of order $k$ and $m$ respectively) play an important role in nonlinear phenomena and their investigation represents an ongoing topic of research \cite{kon, Cam}.  In this paper, the Hamiltonianity of multidimensional  $1+0$ operators has been investigated. In particular, focusing on the case of dimension equal to two, we computed the conditions for an operator of form \eqref{operator2} to be Hamiltonian (up to three components). In this way, we extended the classification of two components non-homogeneous operators proposed by Casati $et \: al.$ \cite{Casati}, by including the missing degenerate case. Furthermore, we presented a classification of the Hamiltonian non-homogeneous operators of form \eqref{operator2} in the case of three dependent variables $u, \,v$ and $
	w$, considering operators both degenerate and non degenerate.\\
	As a further perspective, it could be of interest to extend this classification to higher dimensions and to a greater number of components. Furthermore, we emphasize that the Hamiltonian operators studied in this paper can play an important role in the investigation of integrability of non-homogeneous systems. For this reason, in a future paper, possible bi-Hamiltonian structures associated with this kind of operators will
	be studied. We also remark that, starting from the classification presented in this paper and using the theory of cotangent coverings as in \cite{Ver3, Casati, keri, Vergallo9}, it is possible to proceed with a full classification of the quasilinear systems admitting an Hamiltonian formulation through the operators listed above.
	
	\section*{Acknowledgments}
	The author thanks the financial support of GNFM of the Istituto Nazionale di Alta Matematica, of the PRIN project MIUR Prin 2022, project code
	1074 2022M9BKBC, Grant No. CUP B53D23009350006;
	 and of PRIN 202248TY47 003, "Modelling complex biOlogical systeMs for biofuEl productioN and sTorAge:
	mathematics meets green industry (MOMENTA) ".


\end{document}